\documentclass[prl,twocolumn,preprintnumbers,amsmath,amssymb,a4paper,superscriptaddress,floatfix]{revtex4-1}

\usepackage{graphicx}

\begin{document}
\sloppy 

\title{Revisiting Imidazolium Based Ionic Liquids: Effect of the Conformation Bias of the [NTf$_{2}$] Anion Studied By Molecular Dynamics Simulations}

\author{Benjamin Golub}
\email{benjamin.golub@uni-rostock.de}
\affiliation{Institut f\"ur Chemie, Abteilung Physikalische und Theoretische Chemie, Universit\"at Rostock, Albert-Einstein-Stra{\ss}e 21, D-18059 Rostock, Germany}

\author{Jan Neumann}
\email{jan.neumann@uni-rostock.de}
\affiliation{Institut f\"ur Chemie, Abteilung Physikalische und Theoretische Chemie, Universit\"at Rostock, Albert-Einstein-Stra{\ss}e 21, D-18059 Rostock, Germany}

\author{Lisa-Marie Odebrecht}
\email{lisa-marie.odebrecht@uni-rostock.de}
\affiliation{Institut f\"ur Chemie, Abteilung Physikalische und Theoretische Chemie, Universit\"at Rostock, Albert-Einstein-Stra{\ss}e 21, D-18059 Rostock, Germany}

\author{Ralf Ludwig} 
\email{ralf.ludwig@uni-rostock.de}
\affiliation{Institut f\"ur Chemie, Abteilung Physikalische und Theoretische Chemie, Universit\"at Rostock, Dr.-Lorenz-Weg 2, D-18059 Rostock, Germany}
\affiliation{Leibniz Institut f\"ur Katalyse an der Universit\"at Rostock,
Albert-Einstein-Stra{\ss}e 29a, D-18059 Rostock, Germany}

\author{Dietmar Paschek} 
\email{dietmar.paschek@uni-rostock.de}
\affiliation{Institut f\"ur Chemie, Abteilung Physikalische und Theoretische Chemie, Universit\"at Rostock, Albert-Einstein-Stra{\ss}e 21, D-18059 Rostock, Germany}

\date{\today}

\begin{abstract}
We study ionic liquids composed 1-alkyl-3-methylimidazolium cations
and bis(trifluoromethyl-sulfonyl)imide anions ([C$_n$MIm][NTf$_2$]) with varying chain-length $n\!=\!2, 4, 6, 8$ by using molecular dynamics simulations.
We show that a reparametrization of the dihedral potentials as well as charges of the [NTf$_2$] anion leads to an improvment of the
force field model introduced by K\"oddermann {\em et al.} 
[ChemPhysChem, \textbf{8}, 2464 (2007)]  
(KPL-force field). 
A crucial advantage of the new parameter set is that the minimum 
energy conformations 
of the anion ({\em trans} and
{\em gauche}), as deduced from {\em ab initio} calculations and {\sc Raman} experiments, are now
both well represented by our model.
In addition, the results for  [C$_n$MIm][NTf$_2$] show that this modification
leads to an even better agreement between experiment and molecular dynamics
simulation as demonstrated for densities, diffusion coefficients, 
vaporization enthalpies, reorientational correlation times, 
and viscosities. Even though we  focused 
on a better representation of the anion conformation,
also the alkyl chain-length dependence of the cation behaves
closer to the experiment.
We strongly encourage to use the new NGKPL force field for the [NTf$_2$] anion instead 
of the earlier KPL parameter set for computer simulations aiming to
describe the thermodynamics, dynamics and also structure of imidazolium based ionic liquids.
\end{abstract}

\maketitle

\section{Introduction}

Having a reliable force field available is one of the most important 
prerequisites for setting up a molecular dynamics simulation. 
Hence, a lot of effort has been put
into the development of new as well as the 
improvement of existing force field models. 
There are essentially two different approaches on how to improve or optimize force
fields:

One approach is trying to develop a
``universal'' force field parameter set
which can be applied to a broad range of different molecules or ions, 
such as the force field parameters for ionic liquids introduced
by  P\'adua et
al.\cite{Padua:2004_1,Padua:2004_2,Padua:2006_1,Padua:2006_2,Shimizu:2008,Ishiguro:2008_1,Ishiguro:2008_2,Padua:2009,Lopes:2010,Padua:2012}. 
These force fields are very popular in the ionic liquids
molecular simulation community 
and yield in general 
good results in comparison with experimental data.

An alternative, less universal approach is to focus on 
a specific subset of molecules and ions, and to enhance
the quality of the model by
fitting the parameters of a system to a set of selected
thermodynamical, dynamical and structural properties, which then can be 
accurately emulated by the force field. 
The most well-known example for 
the application of such
a strategy is perhaps the water molecule. 
In 2002 Bertrand Guillot gave a comprehensive overview over  (at the time) more than 40
different water models \cite{Guillot:2002}, 
and the number has been increasing since then
\cite{Baranyai:2012,Bettens:2015,Sadus:2015,Paesani:2016}.
Obviously, water is of great scientific interest. 
As a consequence,
there exist a variety of force field models
 consisting mostly of three (SPC, TIP3P) to five (TIP5P, ST2) 
 interaction sites, including (POL5) or without (SPC/E) polarizability 
an and even force fields  optimized to best represent the 
solid phases of water (TIP4P/ICE) and their phase transitions.

The second strategy was employed by K\"oddermann {\em et al.} in 2007 
to arrive at the KPL (K\"oddermann, Paschek, Ludwig) force field for 
a selected class of imidazolium based ionic liquids composed of
1-alkyl-3-methylimidazolium cations
and bis(trifluoromethyl-sulfonyl)imide anions ([C$_n$MIm][NTf$_2$])
\cite{Koeddermann:2007}. Aim of this
work was to further optimize the force field of P\'adua {\em et al.} to better represent dynamical properties like self-diffusion coefficients, 
reorientational correlation times, and viscosities. As
shown in their original work from 2007 as well as in further works 
published by different groups, the KPL force field has been proven to 
yield reliable results for
dynamical properties, but also thermodynamical properties, such as
the free energies of solvation for light gases in ionic liquids \cite{Kerle:2009,Kerle:2017},
and is still used 
frequently to this date \cite{Daly:2017,Stone:2017,Kazemi:2017}.

\begin{figure}
	\centering
	\includegraphics[width=3.0cm]{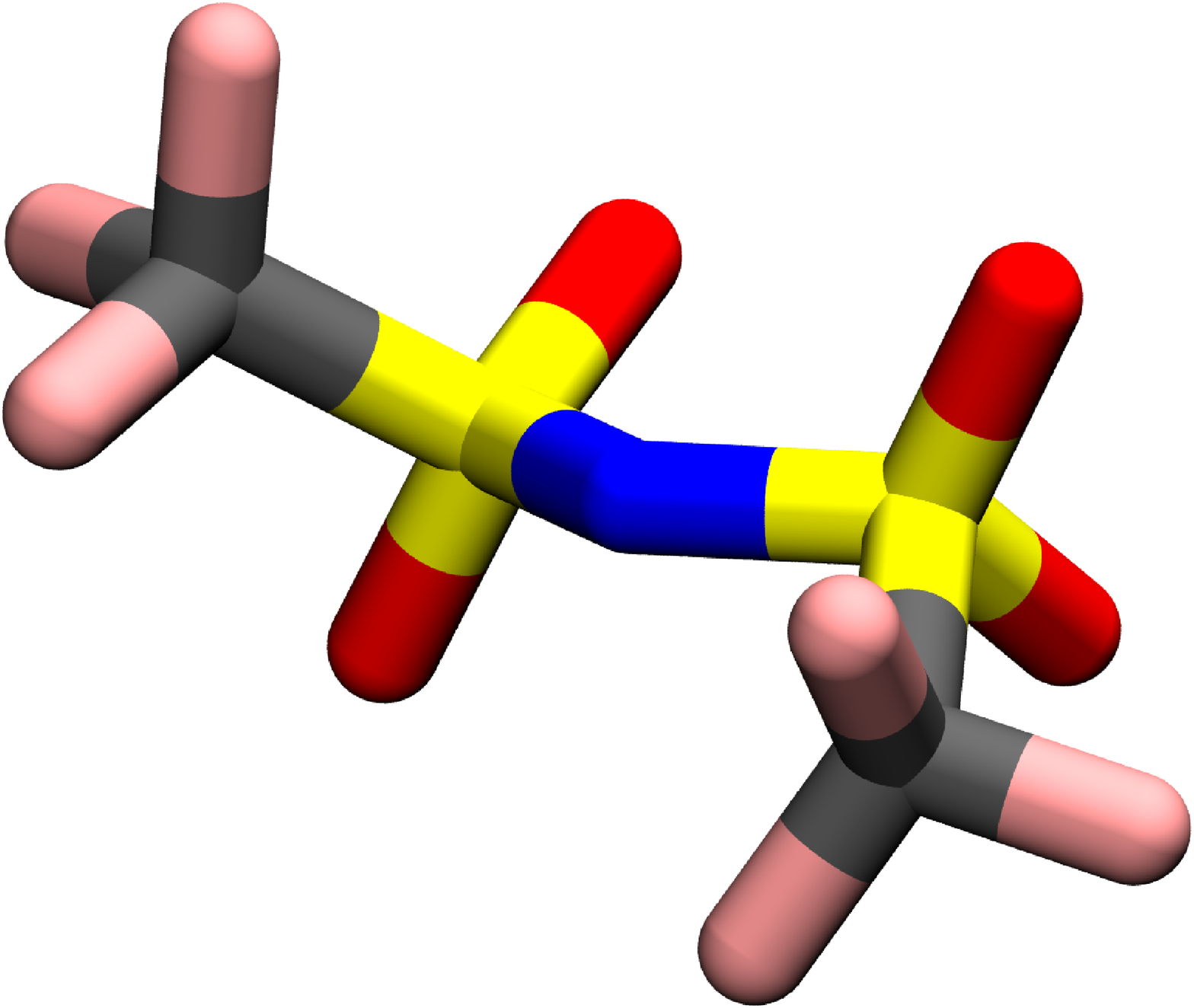} \hspace{0.4cm}
	\includegraphics[width=3.0cm]{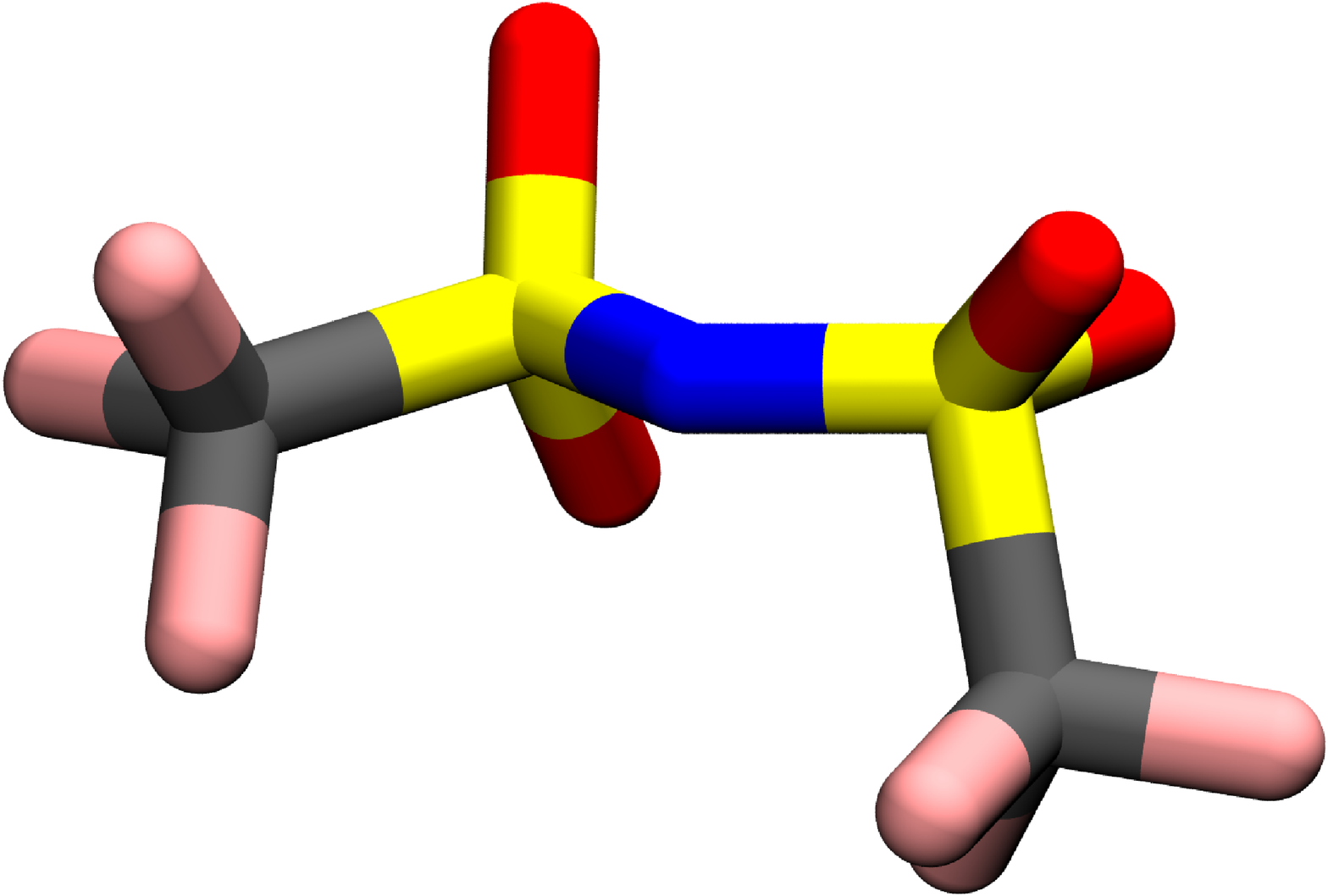}
	\caption{Minimum energy conformations of the [NTf$_2$] anion
	 taken from a MD simulation employing the KPL force field.}
	\label{fig:kodd_conf}
\end{figure}
\begin{figure}
	\centering
	\includegraphics[width=2.5cm]{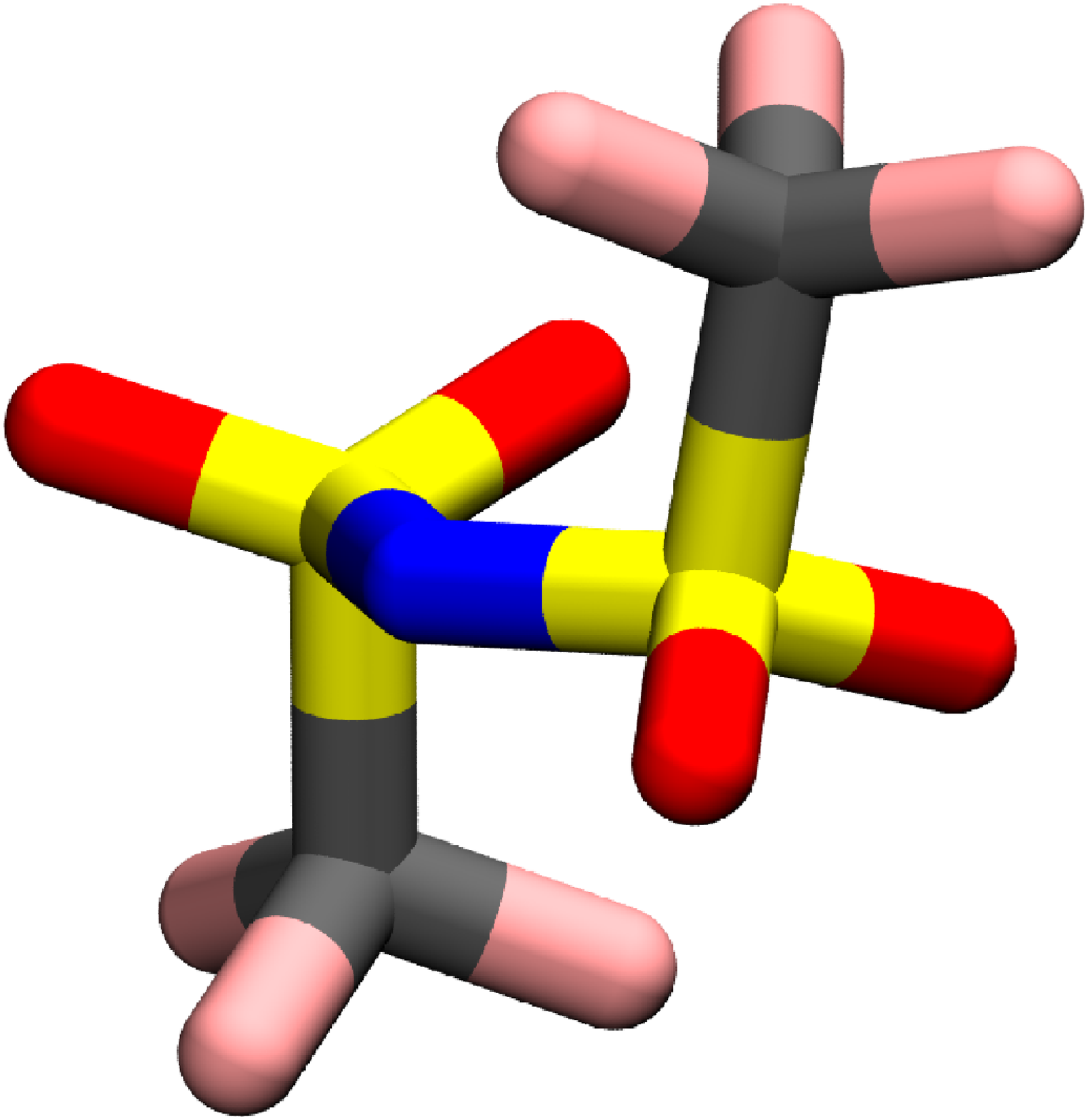} \hspace{0.4cm}
	\includegraphics[width=2.5cm]{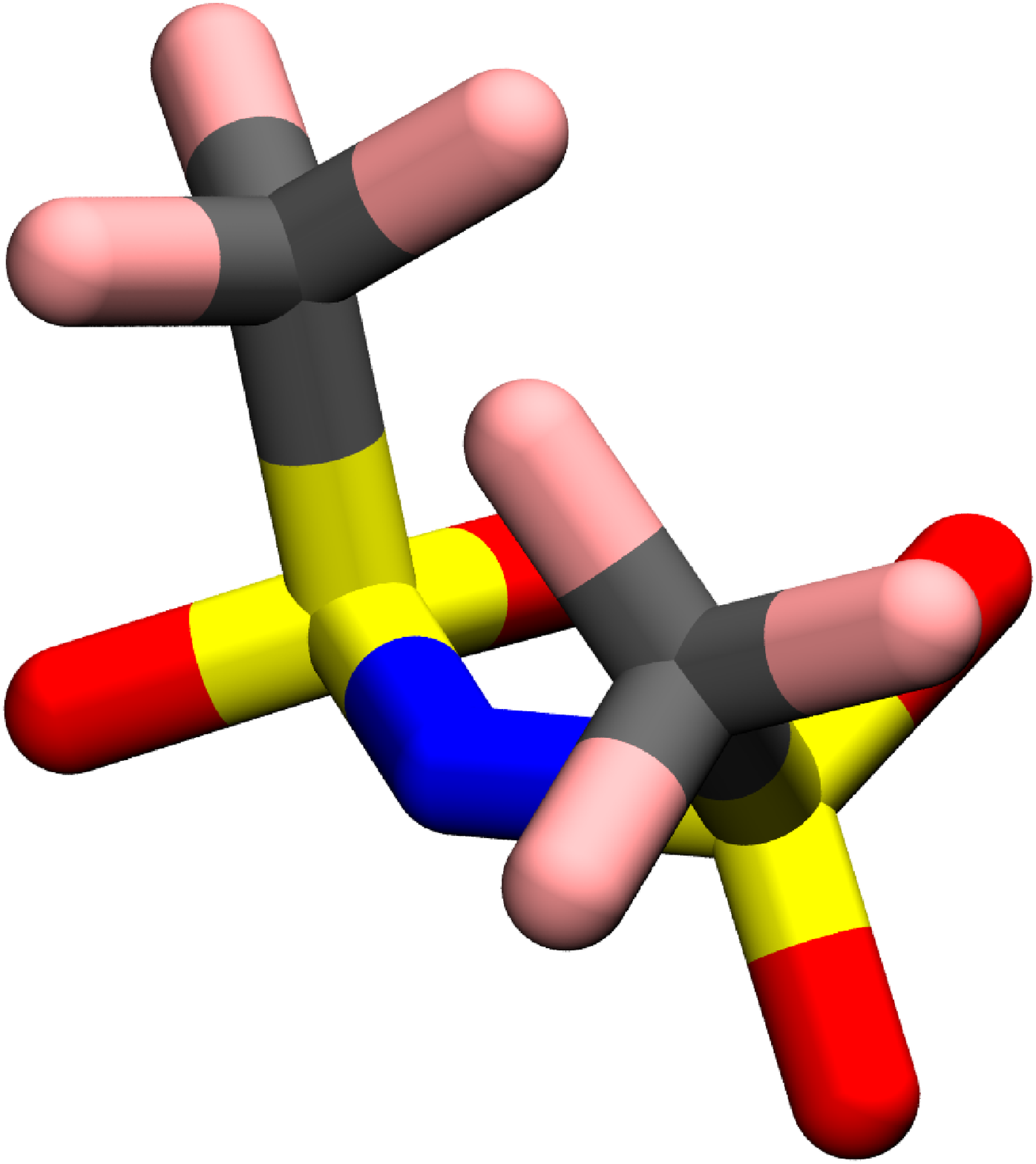}
	\caption{Minimum energy conformations of the [NTf$_2$] anion obtained 
	from {\em ab initio} calculations. The {\em trans} conformation (left) represents
	 the global energy minimum, while the energy of the {\em gauche} conformation 
	 (right) is elevated by about 
	$3\,\mbox{kJ}\,\mbox{mol}^{-1}$.}
	\label{fig:minimum_conf}
\end{figure}
Here we want to present our take on further 
improving the KPL force field by revisiting the 
conformation-space explored by the [NTf$_2$] anion. 
Extensive studies of the conformation of the [NTf$_2$] anion 
using the KPL force field in comparison to experimental as well as
quantum chemical calculations have revealed
a significant mismatch of the energetically favored conformations. 
Therefore we feel the need for presenting
 a modified version of the force field, removing
this conformation-bias. We are discussing the implications of
this modification for
 a wealth of thermodynamical, 
dynamical, and structural quantities.

\section{Conformation-space of the Anion}

During MD simulations of ionic liquids of the type [C$_n$MIm][NTf$_2$]
with the force field of K\"oddermann {\em et al.} 
it became apparent that the favored [NTf$_2$] anion conformations 
observed in the
simulation differ from what has been shown earlier
from quantum chemical calculations \cite{Ishiguro:2008_1}
 as well as from {\sc Raman} experiments
\cite{Fujii:2006} (see Fig. \ref{fig:kodd_conf} and Fig. \ref{fig:minimum_conf}).

\begin{figure}
	\centering
	\includegraphics[width=7.0cm]{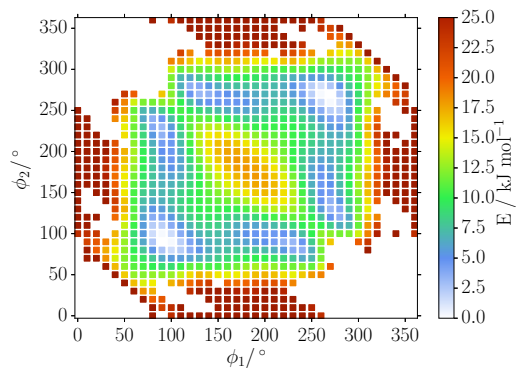}
	\caption{{\em Ab initio} computation of the energy surface of the [NTf$_{2}$] anion as a function of the S1-N-S2-C2, and S2-N-S1-C1  dihedral angles $\phi_{1}$ and $\phi_{2}$. }
	\label{fig:abinitio}
\end{figure}

For locating the minimum energy conformations we performed extensive
 quantum chemical calculations with the {\sc Gaussian 09} 
program \cite{Gaussian_09} 
following the 
approach of P\'adua {\em et al.} \cite{Padua:2004_2}. 
We started by calculating the potential energy surface as a function of the two dihedral angles S1-N-S2-C2
($\phi_{1}$) and S2-N-S1-C1 ($\phi_{2}$) on 
the HF level with a small basis set (6-31G*). 
Subsequent to these optimizations we performed 
single point calculations on the MP2 level 
using the cc-pvtz basis set for all HF optimized conformations. 
In agreement with earlier calculations by P\'adua {\em et al.}
\cite{Ishiguro:2008_1}, and {\sc Raman} measurements of Fujii {\em et al.} \cite{Fujii:2006},
 we observe essentially
 two structurally distinct minimum energy conformations that can be identified
 as energy minima on the energy-landscape
depicted in  Fig. \ref{fig:abinitio}. 
The {\em trans} conformations of the [NTf$_2$] anion are energetically 
preferred, followed by the gauche-conformations,
which are elevated by about $3\,\mbox{kJ}\,\mbox{mol}^{-1}$ 
(see Fig. \ref{fig:minimum_conf}).

\begin{figure}
	\centering
	\includegraphics[width=7.0cm]{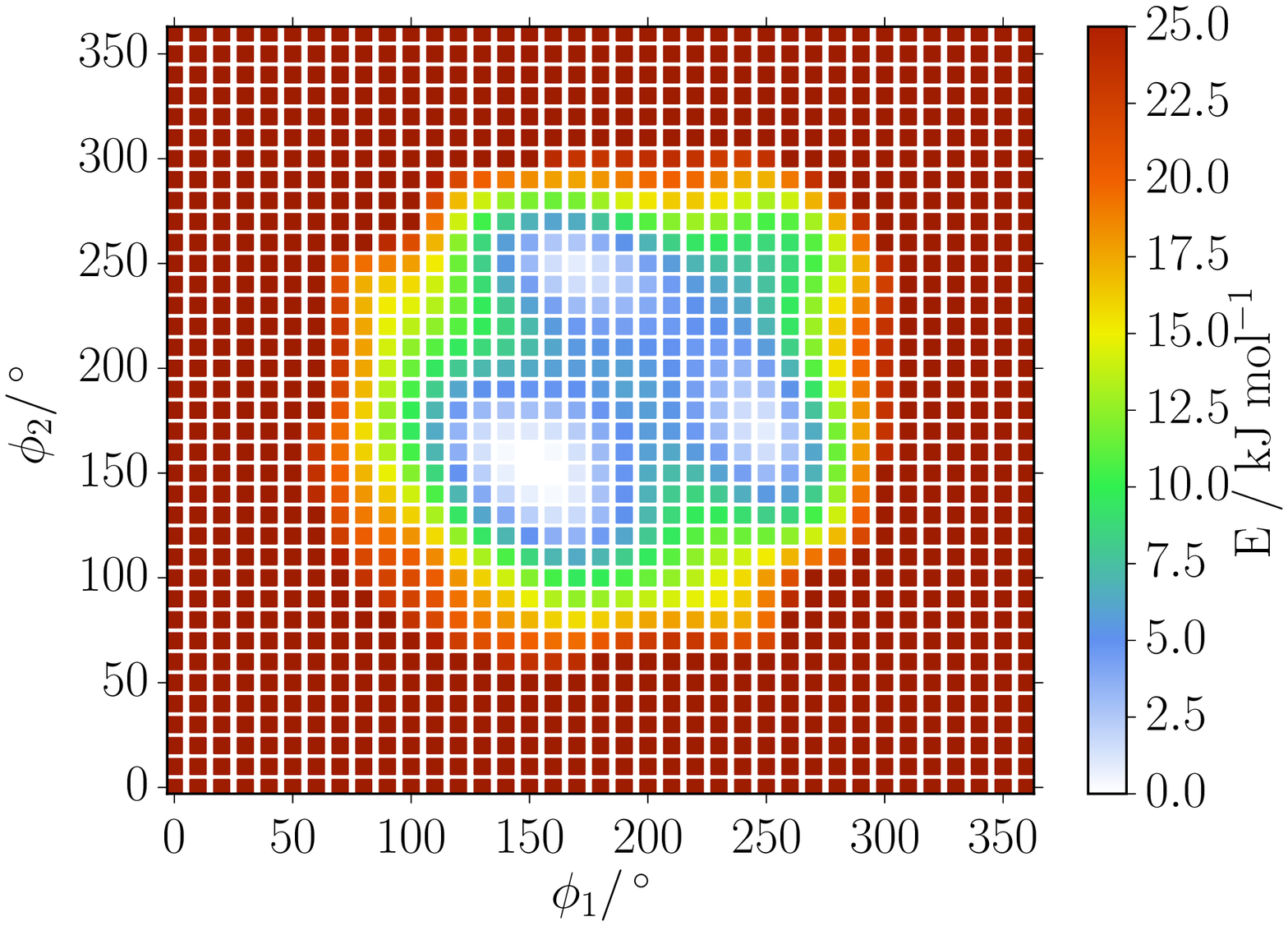}
	\includegraphics[width=7.0cm]{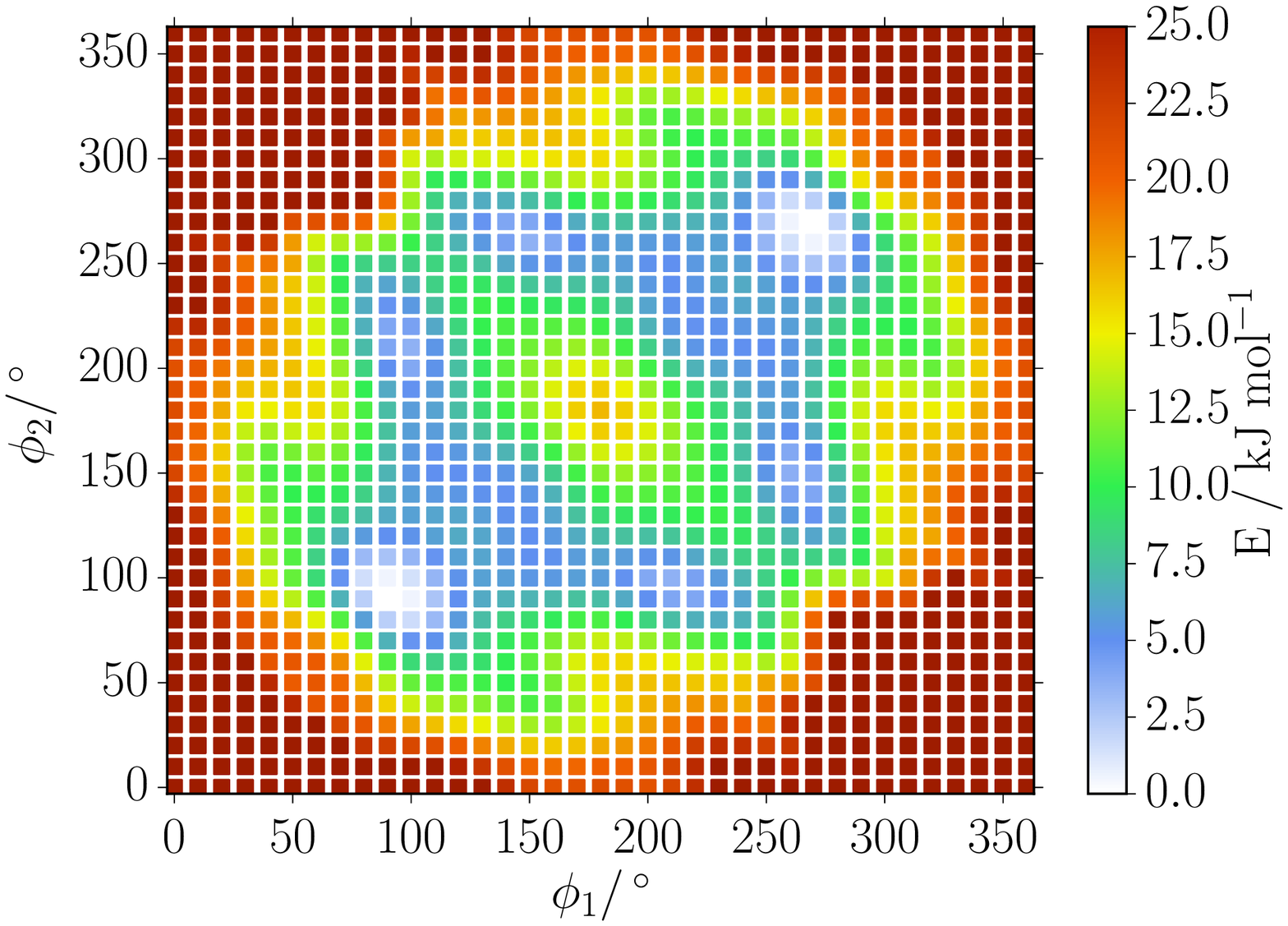}
	\caption{Potential energy surface of the [NTf$_{2}$] anion computed for the
	two force field models. The new force field (bottom panel) provides a much better
	representation of the {\em ab initio} calculations shown in Fig. \ref{fig:abinitio} than the
	 the original KPL force field (top panel).} 
	\label{fig:dihe_sim}
\end{figure}

To compare these {\em ab initio} calculations with 
the KPL force field model, we employed the molecular 
dynamics package {\sc Moscito} 4.180 and computed the
same potential energy surface as a function of the two
dihedral angles $\phi_{1}$ and $\phi_{2}$ (see Fig. \ref{fig:dihe_sim} top panel)
by fixing the two dihedral angles and optimizing all other degrees of freedom.
We would like to add that in the force field-optimizations
all bond-lengths were kept fixed. 
It is quite obvious that the KPL force field does not adequately
 reproduce the potential energy surface obtained from the quantum
chemical calculations
(compare the top panel in Fig. \ref{fig:dihe_sim} with FIG \ref{fig:abinitio}).
The minimum energy conformations of the KPL model reveals essentially
two structurally distinct conformations illustrated in
Fig. \ref{fig:kodd_conf}. However, both are somewhat similar, being 
positioned between the {\em trans} and {\em gauche }conformations
favoured in the {\em ab initio} calculations.
The fact that the
energy landscape does not reflect all the 
symmetry-features of the molecule, however,
might be a lesser problem since energy barriers are rather large and
the anion could explore similar conformations simply by rotation.

\begin{figure}
	\centering
	\includegraphics[width=7.0cm]{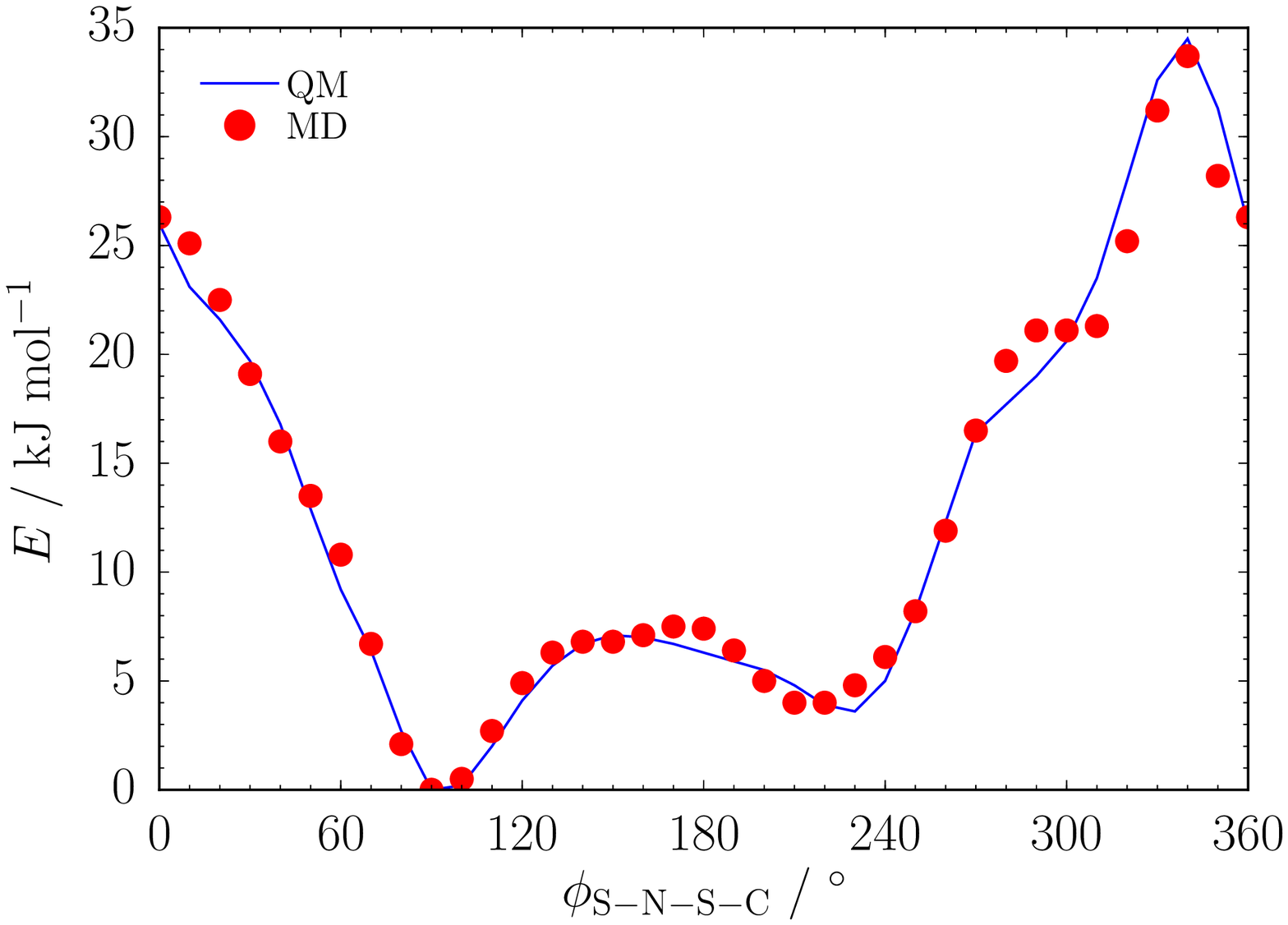}
	\includegraphics[width=7.0cm]{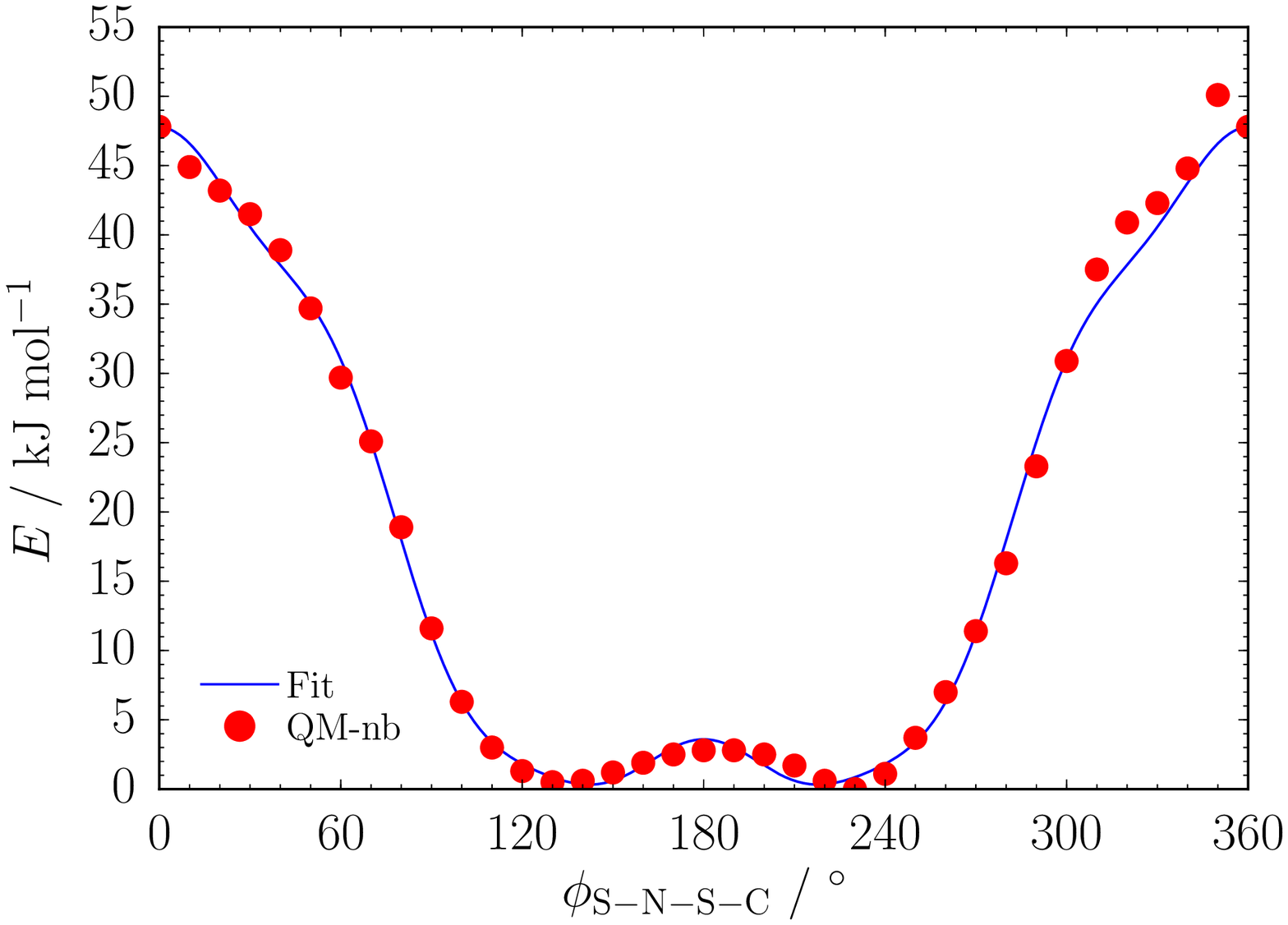}
	\caption{Top panel: Potential energy of the entire [NTf$_2$] anion as a function of the
	 of S2-N-S1-C1 dihedral angle $\phi_{2}$ with $\phi_{1}$ being fixed at 
	 $\phi_{1}\!=\!90^\circ$. Bottom panel: Torsion potential fitted to the difference between QM and force field model (with switched off torsion potential).}
	\label{fig:snsc_pot}
\end{figure}

However, for arriving at a better representation of the {\em ab initio} energy surface,
we reparameterized the charges 
as well as the two distinct independent dihedral potentials (S-N-S-C and
F-C-S-N), while keeping the other parameters unchanged. 
From our quantum chemical calculations we yield the 
global minimum conformations at
$\phi_{1}\!=\!\phi_{2}\!=\!90^\circ$ and $\phi_{1}\!=\!\phi_{2}\!=\!270^\circ$. Due to the symmetry of the [NTf$_{2}$] anion these two minima 
are conformationally identical. 
To calculate the parameters for the S-N-S-C dihedral angle,
 we fixed $\phi_{1}$ at $90^\circ$
and calculated the energy as function of the dihedral angle $\phi_{2}$ on 
the MP2 level using a cc-pvtz basis set 
(as shown in Fig. \ref{fig:snsc_pot}). 
The same procedure was applied using the KPL force field
while switching of the dihedral potential,
such that  only the
nonbonding (nb) interactions matter. We then subtracted 
the latter energy function from the energies obtained via 
the QM calculations, and arrive at the dihedral potential for
the dihedral angle S-N-S-C, 
which should be reproduced by the torsion potential in our force field 
(see Fig. \ref{fig:snsc_pot} bottom panel).

\begin{figure}
	\centering
	\includegraphics[width=7.0cm]{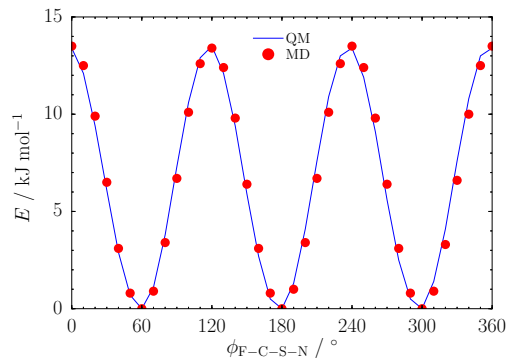}
	\caption{Potential energy of the [NTf$_2$] anion as function of the
	 F-C-S-N dihedral angle.}
	\label{fig:fcsn_pot}
\end{figure}

In contrast to K\"oddermann {\em et al.}, we chose to fit a dihedral potential function
obeying the conformational symmetry-features of the anion using
\begin{equation}
V_{\kappa \lambda \omega \tau}^{\rm dp}=\sum_{n}k_{m}^{\rm dp}[1 + \cos({m_{n}\psi_{m}-\psi^{0}_{m})] }
\end{equation}
(with $n\!=\!6$ and $\psi^{0}_{m}\!=\!0$) to the computed {\em ab initio} potential, leading to
the proper minimum energy conformations of the [NTf$_2$] anion \cite{Padua:2004_2}.
Similarly obtained were parameters for the F-C-S-N dihedral potential
of the terminal $\mbox{C}\mbox{F}_3$-groups
(see Fig. \ref{fig:fcsn_pot}). The complete set of 
new parameters for the NGKPL force field is given in
Table \ref{tab:ntf2_dieder}. All charges were computed
from the MP2-wavefunction using the method of Merz and Kollman 
as implemented in in the {\sc Gaussian 09} programm.
\cite{Singh:1984}. The refined charges are 
 listed in Table \ref{tab:ntf2_ljq}.

Finally, employing new refined parameters for the dihedral potentials and 
partial charges, we re-calculated the energy surface as a function of 
the two dihedral angles $\phi_{1}$ and
$\phi_{2}$ (see Fig. \ref{fig:dihe_sim} bottom panel). The result is 
in much better agreement with the {\em ab initio} calculations
and resolves the conformational mismatch issue for 
the force field of the [NTf$_2$] anion.

All parameters for the new [NTf$_2$] anion force field are listed in the Tables \ref{tab:ntf2_ljq}-\ref{tab:ntf2_dieder}. The
original parameters as well as the parameters for the cations can be found in the publication of K\"oddermann {\em et al.}
\cite{Koeddermann:2007}.

\begin{table}
	\centering
     \setlength{\tabcolsep}{0.7cm}
         \caption{{\sc Lennard-Jones} parameters $\sigma$, $\epsilon$ and
charges $q$ for all interaction sites of the [NTf${_2}$] anion.}
     \begin{tabular}{crrr}\hline\hline 
site & $\sigma$ / \AA & $\epsilon$ / K & $q$ / e\\
     \hline
    F &  2.6550  &   8.00 & -0.189 \\
    C &  3.1500  &   9.96 &  0.494 \\
    S &  4.0825 &  37.73 &  1.076 \\
    O &  3.4632 &  31.70 & -0.579 \\
    N &  3.2500 &  25.66 & -0.690 \\ \hline\hline
     \end{tabular}
     \label{tab:ntf2_ljq}
\end{table}

\begin{table}
	\centering
    \setlength{\tabcolsep}{0.15cm}
             \caption{Bond length $r^0_{\kappa \lambda}$ and angle
parameters $\phi^{0}_{\kappa
\lambda \omega}$ and $k^{\rm a}_{\kappa \lambda \omega}$ for the angle
potential $V_{\kappa \lambda \omega}^{\rm a}\!=\! \tfrac{1}{2}
k^{\rm a}_{\kappa \lambda \omega}(\phi_{\kappa \lambda \omega} -
\phi^{0}_{\kappa \lambda \omega})^{2}$  in the force field of the
[NTf$_2$] anions. }
                 \begin{tabular}{ccccc}\hline\hline 
     bond  & $r^0_{\kappa \lambda}$ / \AA       &   angle & $\phi^0_{\kappa
\lambda \omega}$ / $^\circ$   & $k^{\rm a}_{\kappa \lambda \omega}$ / kJ
mol$^{-1}$rad$^{-2}$ \\
     \hline
     C-F & 1.323  &   F-C-F   & 107.1 & 781.0 \\
     C-S & 1.818  &   S-C-F   & 111.8 & 694.0 \\
     S-O & 1.442  &   C-S-O   & 102.6 & 870.0 \\
     N-S & 1.570  &   O-S-O   & 118.5 & 969.0 \\
         &              &   O-S-N   & 113.6 & 789.0 \\
         &              &   C-S-N   & 100.2 & 816.0 \\
         &             &   S-N-S   & 125.6 & 671.0 \\ \hline\hline
         \end{tabular}
         \label{tab:ntf2_bond}
\end{table}

\begin{table}
    \centering
    \setlength{\tabcolsep}{0.26cm}
     \caption{Parameters $k^{\rm dp}_{m}$ and $\psi^{0}_{m}$ for the torsion
potential $V_{\kappa \lambda \omega \tau}^{\rm dp}
=
\sum_{n}k_{m}^{\rm dp}[1 + \cos({m_{n}\psi_{m}-\psi^{0}_{m})] }$ in the
force field of the [NTf$_2$] anion.}
\begin{tabular}{cccrc}\hline\hline 
  & $n(\kappa \lambda \omega \tau)$ &  $m_{n}$ & $k^{\rm dp}_{m}$ / kJ
mol$^{-1}$ & $\psi^{0}_{m}$ / $^\circ$ \\
   \hline
  F-C-S-N &1 & 3        & 2.0401  & 0.0 \\
   \hline
  S-N-S-C &1 & 1  &  23.7647   & 0.0 \\
  &2 & 2   &   6.2081   & 0.0 \\
  &3 & 3   &  -2.3684   & 0.0 \\
  &4 & 4   &  -0.0298   & 0.0 \\
& 5 & 5   &   0.6905   & 0.0 \\
& 6 & 6   &   1.0165   & 0.0 \\\hline\hline
   \end{tabular}
   \label{tab:ntf2_dieder}
\end{table}

\section{Molecular Dynamics Simulations}

We performed MD simulations for the two force fields KPL and NGKPL with {\sc Gromacs 5.0.6} \cite{Lindahl:2001, Berendsen:1995,
Hess:2008, Pronk:2013, Gromacs_5.0.6} over a temperature range from
$T\!=\!273$ -- $483\,\mbox{K}$ to calculate thermodynamical and dynamical 
properties and compare them with the original KPL force field. All simulations 
were carried out in the $NpT$ ensemble.
However, to compute viscosities, we performed additional $NVT$ simulations using
starting configurations sampled along the $NpT$-trajectory.
Periodic boundary conditions where applied using cubic simulation boxes
containing 512 ion-pairs. 
We applied smooth particle mesh {\sc Ewald} summation \cite{Essmann:1995} for the electrostatic interactions with a real space cutoff of
$0.9\,\mbox{nm}$, a mesh spacing of $0.12\,\mbox{nm}$ and 4th order interpolation. The {\sc Ewald} convergence factor $\alpha$ was set to
$3.38\,\mbox{nm$^{-1}$}$ (corresponding to a relative accuracy of the {\sc Ewald} sum of $10^{-5}$). All simulationa were 
carried out with a timestep of $2.0\,\mbox{fs}$, while 
keeping bond lengths fixed using the LINCS algorithm \cite{Hess:1997}.

An initial equilibration was done for $2\,\mbox{ns}$ at $T\!=\!500\,\mbox{K}$ applying {\sc Berendsen} thermostat as well as {\sc Berendsen}
barostat with coupling times $\tau_\textrm{T}\!=\!\tau_\textrm{p}\!=\!0.5\,\mbox{ps}$ \cite{Berendsen:1984}. After this another equilibration was done for
$2\,\mbox{ns}$ at each of the desired temperatures. For each of the six temperatures 273\,K, 303\,K, 343\,K, 383\,K, 423\,K and 483\,K
we performed production runs of $30\,\mbox{ns}$, keeping the
the pressure 
fixed 
at $1\,\mbox{bar}$ applying {\sc Nos\'e-Hoover} thermostats
\cite{Nose:1984,Hoover:1985} with $\tau_\textrm{T}\!=\!1\,\mbox{ps}$ and {\sc Rahman-Parrinello} barostats \cite{Parrinello:1981,Nose:1983} with
$\tau_\textrm{p}\!=\!2\,\mbox{ps}$. 

\section{Results \& Discussion}

Analogous to the publication of 
K\"oddermann {\em et al.} from 2007 \cite{Koeddermann:2007} 
we will compare densities, self-diffusion coefficients
and vaporization enthalpies for [C$_n$MIm][NTf$_2$]
 as function of temperature and alkyl chain-length 
as well as viscosities and
reorientational correlation times for [C$_2$MIm][NTf$_2$] as function of 
temperature. It is important to keep in mind, that the original
force field was optimized to reproduce these properties and yields a good agreement between experiment and simulation. By resolving the
mismatch of the favored conformations of the [NTf$_2$] 
anion we are able to describe these properties as 
good as the KPL force field or even
better.

\subsection{Structural Features}

\begin{figure*}
	\includegraphics[width=15cm]{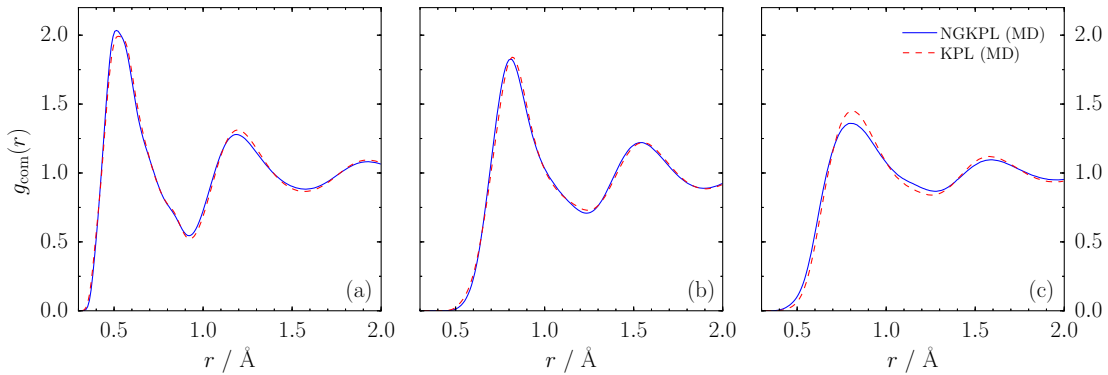}
	\caption{Center of mass radial pair distribution functions for cation-anion (a), anion-anion (b) and cation-cation (c) for [C$_{
2}$MIm][NTf$_2$] at $T=303\,\mbox{K}$. The NGKPL-data are shown as blue lines 
and the KPL-data force as red dashed lines.}
	\label{fig:grcom}
\end{figure*}

\begin{figure}
	\includegraphics[width=7.0cm]{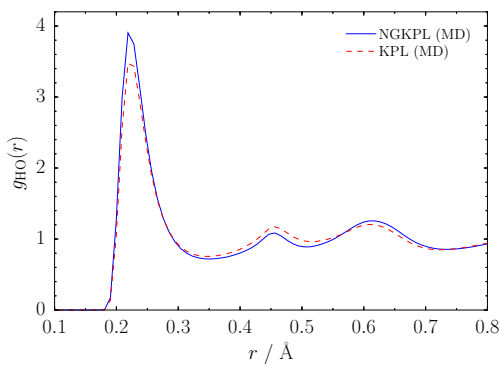}
	\caption{Radial pair distribution function of the anion-oxygens around the C(2) hydrogen site on the cation for [C$_2$MIm][NTf$_2$] at
$T=303\,\mbox{K}$. The NGKPL force field is shown in the blue line and the KPL force field in the red dashed line.}
	\label{fig:groh}
\end{figure}

\begin{figure}
	\includegraphics[width=8.0cm]{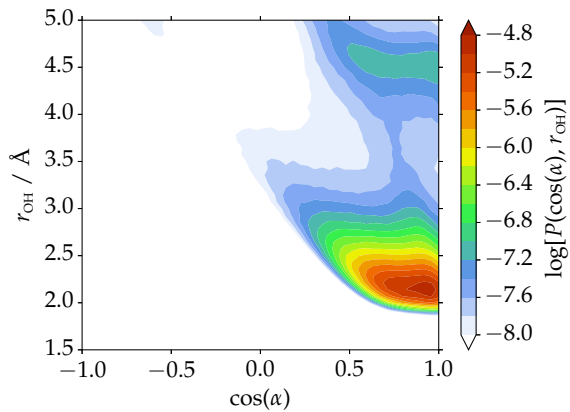}
	\caption{Probability density of the anion-oxygens around the
	C(2) hydrogen sites
	as function of the intermolecular
distance $r_\textrm{OH}$ and the angle between the C(2)-H bond-vector and 
the intermolecular C(cation)-O(anion) vector. Shown for the NGKPL force field at
$T=303\,\mbox{K}$.}
	\label{fig:histo}
\end{figure}

Here we take a look at structural features of the liquid phase
and in how they are influenced by
changes in the conformation-population of the [NTF$_2$] anion. First we
inspect the three distinct 
center of mass pair distribution functions between the different ions
computed for [C$_n$MIm][NTf$_2$] with $n=2$ at
$T\!=\!303\,\mbox{K}$ (shown in Fig. \ref{fig:grcom}).
It is quite apparent that these distribution functions are only 
slightly affected by the alterations in the force field. Most notable are the
differences observed in the anion-anion pair distribution
function depicted in Fig. \ref{fig:grcom}c with the first peak being significantly
broadened. It is quite obvious to assume that this behavior is related to
the more distinct conformational states  ({\em trans} and {\em gauche}) 
that the reparameterized  [NTf$_2$] anion is adopting as
shown in Fig. \ref{fig:minimum_conf}. In the {\em trans} state   
the molecule is more elongated along the molecular axis and more compact
perpendicular to it. In addition, the gauche-state is generally
more compact than the minimum energy conformations adopted
by the original KPL force field model shown in Fig. \ref{fig:kodd_conf}. This leads to an enhanced population of
both, short and long anion-anion distances.
This effect  manifests itself also in the
slight shift of the maximum of the first peak of the anion-cation pair distribution
function towards smaller distances (see Fig. \ref{fig:grcom}a).
Another interesting distribution function 
is the pair distribution function of the anion-oxygens 
surrounding the C(2)-hydrogen site
on the cation. The C(2)-position is deemed to act as a 
hydrogen-bond donor \cite{Fumino:2008, Wulf:2010}.
With changing conformations we expect an effect on 
 the hydrogen bonding situation between the anion and
cation. 
Here we observe that the NGKPL force field promotes hydrogen bonds between 
anions and cations as indicated by
an increased first peak of the
O-H pair distribution function shown in Fig. \ref{fig:groh}. 
The computed number of hydrogen bonds increases
 throughout by about $4\,\%$, mostly unaffected by the alkyl chain-length
 and temperature (not shown). 
Taking into account the importance of more elongated {\em trans} configurations
of the anion, it is also not surprising that
the second peak is somewhat depleted, while 
the third peak is again enhanced (see Fig. \ref{fig:groh}).
We further investigate the hydrogen-bond situation by not just
 looking at the distance between the oxygen and hydrogen, but also
at the angular distribution. 
Therefore we compute the probability density map
of the anion-oxygens
surrounding the  C(2) hydrogen site on the cation. Again we focus
on the C(2) hydrogen because its hydrogen-bond interaction with the anion is
deemed the strongest and most important. To calculate this  map 
we compute both, the O-H distance as
well as  angle between the C-H bond-vector on the cation and the intermolecular C-O vector,
where C is the C(2)-position of the cation and O represents the oxygen-sites on the anions. 
In addition, the computed probabilities are weighted by $r^{-2}_\textrm{OH}$.  
It is revealed that the maximum 
of this probability density map does not quite represent 
a linear hydrogen bond at a distance of $2.3\,\mbox{\AA}$, but is tilted by about
$25^\circ$, and is characterized by a rather broad angular distribution. 
(Fig. \ref{fig:histo}).

\subsection{Densities \& Self-Diffusion Coefficients}

\begin{figure}
	\includegraphics[width=7.0cm]{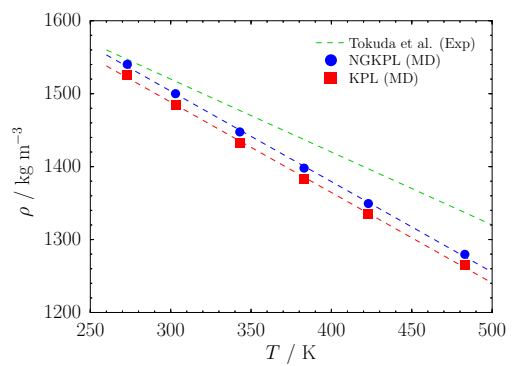}
	\caption{Mass densities of [C$_2$MIm][NTf$_2$] as function of temperature.
The experimental data of 
Tokuda {\em et al.} is given according to their fitted temperature dependence
 (green dashed line) \cite{Tokuda:2005}. The
results from our molecular dynamics simulation 
using the NGKPL (blue dots) and KPL (red squares)
force fields were fitted with a linear function represented by
the dashed lines. See also Table \ref{tab:density_diffu_c2mim}.}
	\label{fig:density}
\end{figure}

To get an idea on how the changing conformation-populations
influence the properties of the imidazolium based ionic liquids, we 
first take a look at the mass density of [C$_2$MIm][NTf$_2$]. 
In molecular simulations the density has always been an important property 
for evaluating a force field. 
The enhanced conformational diversity of the [NTf$_2$] anion 
leads to a slight increase in the density over the whole temperature range 
(see Fig.\ref{fig:density}). This overall increase is in better
agreement with the experimental data from Tokuda {\em et al.} \cite{Tokuda:2005}. For lower
temperatures the NGKPL force field even matches the experimental values. 
The thermal expansivity, however, is significantly overestimated, although
at the highest temperatures the difference between experiment and
simulation is still within about $5\;\%$.
Despite the overall density increase from KPL to NGKPL, the 
thermal expansivities of both models are practically identical.

\begin{figure}
	\includegraphics[width=7.0cm]{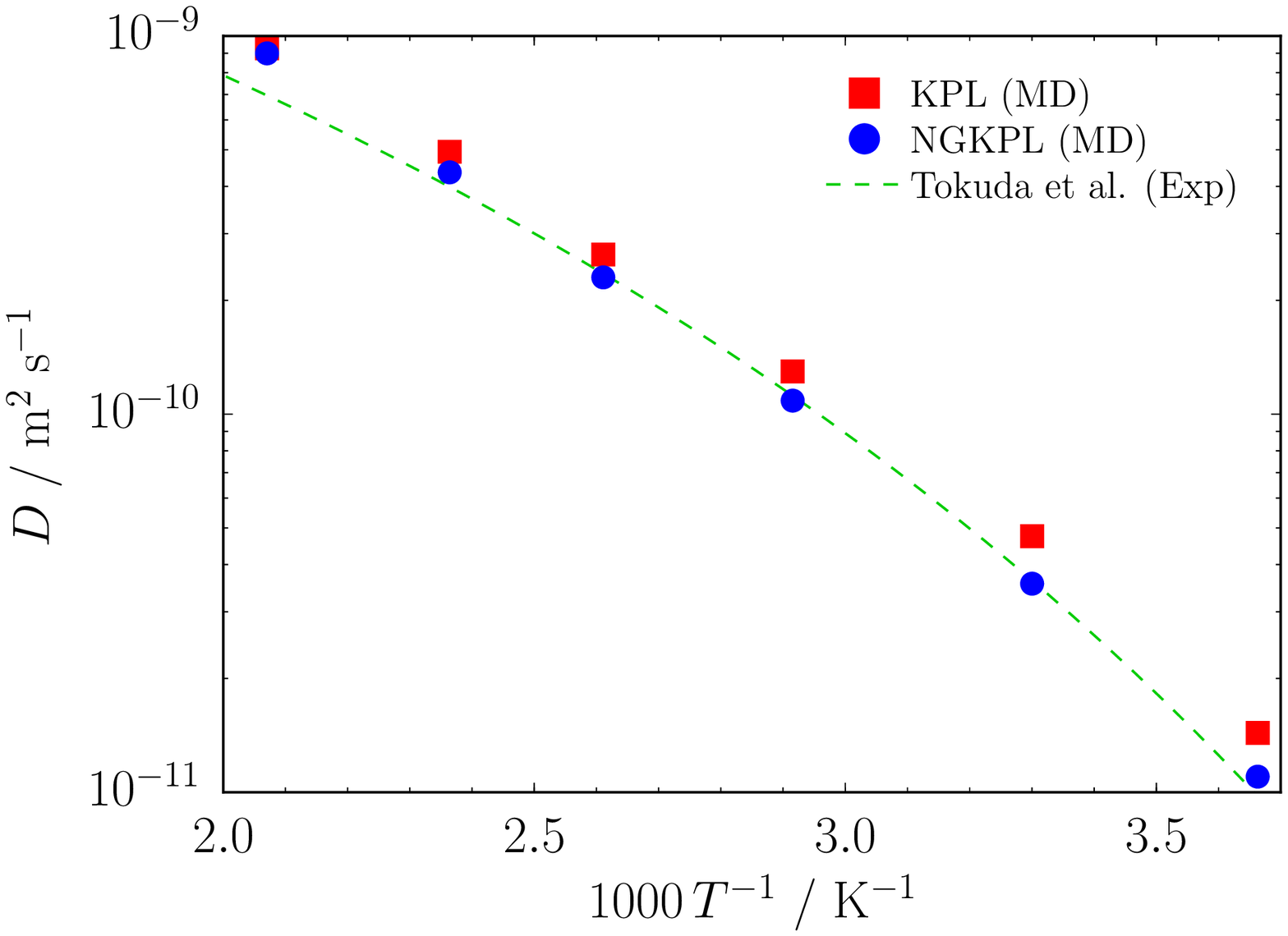}
	\caption{Self-diffusion coefficients as function of the temperature for [C$_{2}$MIm][NTf$_2$]. The
experimental data of Tokuda {\em et al.} are represented 
according to their fitted temperature dependence
(green dashed line) \cite{Tokuda:2005}. The red squares (KPL) and
blue dots (NGKPL) represent the results from our molecular 
dynamics simulations. See also Table \ref{tab:density_diffu_c2mim}.} 
	\label{fig:diffusion}
\end{figure}

\begin{figure}
	\includegraphics[width=7.0cm]{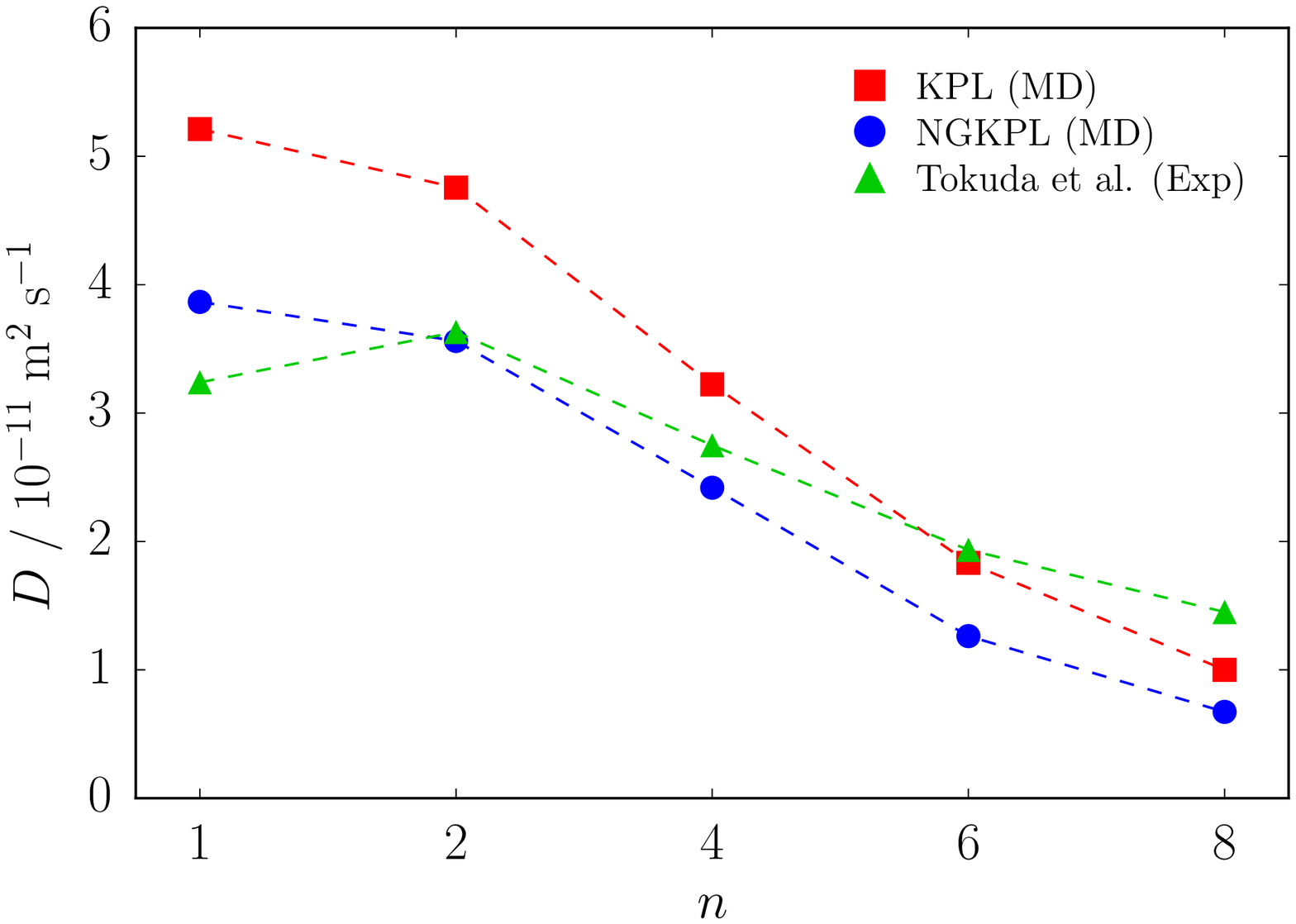}
	\caption{Self-diffusion coefficients as a function of the alkyl chain-length for [C$_n$MIm][NTf$_2$] at $T=303\,\mbox{K}$. The
experimental data are shown as green triangles, the KPL force field as red squares and the NGKPL force field as blue dots. The dashed lines
are only guides for the eye. See also Table \ref{tab:diffu_vap}.}
	\label{fig:diffusion_2}
\end{figure}

With this increasing density, also 
slightly reduced self-diffusion coefficients for the [NTf$_2$] anion 
are observed (see Fig. \ref{fig:diffusion}). 
We calculated the self-diffusion coefficient using the {\sc Einstein} relation
\begin{align}
	D=\frac{1}{6}\lim\limits_{t \to \infty}\frac{d}{dt}\left\langle \left| \vec{r}{_i}(t) - \vec{r}{_i}(0)\right|^2  \right\rangle
\end{align}
as function of the temperature for [C$_2$MIm][NTf$_2$] (Fig. \ref{fig:diffusion}) as well as as function of the alkyl chain-length of [C$_n$MIm][NTf$_2$] at
$T=303\,\mbox{K}$ (Fig. \ref{fig:diffusion_2}).
As shown in 2007 the KPL force field is able to yield self-diffusion coeffients in good agreement with the experimental data. Nevertheless, using
the new NGKPL parameters we are able to describe the temperature dependence of the self-diffusion coefficient of the [NTf$_2$] anion in
[C$_2$MIm][NTf$_2$] even better (Fig. \ref{fig:diffusion}).
Taking a look at the alkyl chain-length dependence we can support the findings for the $n=2$ imidazolium ionic liquid. The NGKPL force
field is able to reproduce the dependence better, especially for $n\leq4$, for 
longer chains the KPL force field is closer to the experiment
(see Fig. \ref{fig:diffusion_2}). As observed for the temperature dependence the general trend of the self-diffusion coefficient as function of
the alkyl chain-length is identical for the KPL and NGKPL force field.
\begin{table}
	\centering
     \setlength{\tabcolsep}{0.39cm}
     \caption{Temperature dependence of the density $\rho$ and the self-diffusion coefficients of the [NTf$_{2}$] anion $D_{-}$ in
[C$_{2}$MIm][NTf$_{2}$] according to the KPL and NGKPL force fields. See also Fig. \ref{fig:density} and Fig. \ref{fig:diffusion}.}
     \begin{tabular}{crrrr}\hline\hline
         & \multicolumn{2}{c}{$\rho$ / kg\,m$^{-3}$} &
\multicolumn{2}{c}{$D_{-}$ / 10$^{-11}$ m$^{2}$\,s$^{-1}$}\\
         $T$ / K & KPL    & NGKPL    &    KPL    & NGKPL \\ \hline
         273 & 1525 & 1540 & 1.4 & 1.1\\
         303 & 1485 & 1500 & 4.8 & 3.6\\
         343 & 1433 & 1448 & 13.0 & 10.9\\
         383 & 1383 & 1398 & 26.4 & 23.0\\
         423 & 1335 & 1349 & 49.4 & 43.6\\
         483 & 1265 & 1280 & 92.7 & 90.0\\\hline\hline
     \end{tabular}
     \label{tab:density_diffu_c2mim}
\end{table}

\subsection{Vaporization Enthalpies}

\begin{figure}
	\includegraphics[width=7.0cm]{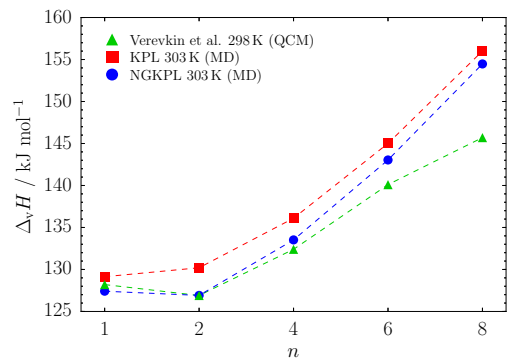}
	\caption{Vaporization enthalpies as function of the alkyl chain-length for NGKPL (blue dots) KPL (red squares) at $T=303\,\mbox{K}$.
For comparison we also show the QCM data of Verevkin {\em et al.} for $T=298\,\mbox{K}$ (green triangles) \cite{Verevkin:2013}. See also Table
\ref{tab:diffu_vap}.}
	\label{fig:vH}
\end{figure}

The magnitude of the vaporization enthalpy of 
ionic liquids 
was studied extensively over the last few years and has been sometimes
discussed quite
emotionally \cite{Zaitsau:2006, Armstrong:2007, Heintz:2007, Santos:2007, Luo:2008, Heym:2011, Rocha:2011, Verevkin:2013, Schroeder:2014}. 
For the purpose of this study we will compare our results with the more
recent QCM data of imidazolium based ILs of type
[C$_n$MIm][NTf$_2$]
from Verevkin {\em et al.} of 2013 \cite{Verevkin:2013} 
as shown in Fig. \ref{fig:vH}.
We would like to point out that an exhaustive
overview of the huge amount of vaporization
enthalpy data 
 from different experiments 
 as well as molecular 
 simulation studies is provided
  in the supporting informations of Verevkin {\em et al.} \cite{Verevkin:2013} 
and in the COSMOS-RS study by Schr\"oder and Coutinho \cite{Schroeder:2014}. 
The vaporization enthalpies per mol
of [C$_n$MIm][NTf$_2$] were here calculated by assuming
ideal gas behavior with
\begin{equation}
	\Delta_{\rm v}H\approx\Delta_{\rm v}U + RT\;,
	\label{eq:dvh1}
\end{equation}
which is a well justified approximation, given the low vapor pressures
of ILs at low temperatures.
The energy difference between the liquid and gas phases were computed via
\begin{equation}
	\Delta_{\rm v}U=U'_{\rm g} - U'_{\rm l}
	\label{eq:dvh2}
\end{equation}
where $U'_{\rm l}$ and $U'_{\rm g}$ are
the internal energies per mol ion-pairs of the liquid and gas phases, respectively.
To determine $U'_{\rm g}$ we performed gas phase simulations 
of individual ion-pairs without periodic boundary conditions. 
It has been shown in the literature that the 
gas phase of ionic liquids consists mostly of ion-pairs
\cite{Zaitsau:2016,Verevkin:2011, Verevkin:2013, Zaitsau:2012, Verevkin:2012, Ahrenberg:2014, Verevkin:2012_2, Boeck:2014} tied together by strong
long-range electrostatic forces.
Hence, simulating an isolated ion-pair instead of
separated ions is the most realistic
approximation of the IL gas phase. 
As it is standard practice, during the simulation of 
of both, the liquid phase and also of the isolated ion-pair,
 the total linear momentum was set to zero, thus
eliminating the systems center of mass
 translational motion. In addition, in the simulations of the
 isolated ion-pairs also the total angular momentum was set to zero.
However, when comparing the internal energy of the gas phase and the liquid phase, 
we have to correct for differences in the kinetic energy stored
in the translational/rotation motion of either system by adding
\begin{eqnarray}
	U'_{\rm g} &=&U_{\rm g} + \frac{6}{2}RT \\  
	U'_{\rm l} &=&U_{\rm l} + \frac{3}{2}RT\times \frac{1}{N_\textrm{IP}} 
\end{eqnarray}
per mole of ion-pairs, where $N_\textrm{IP}=512$ is the number
of ion-pairs used in the liquid simulation,
and $U_{\rm g}$ and $U_{\rm g}$ are the total energies per ion-pair as computed
directly from the MD simulations.
With these corrected molar internal energies $U'_{\rm g}$ and $U'_{\rm l}$ we 
compute the heat of vaporization $\Delta_{\rm v}H$ using Eq. \ref{eq:dvh1}
for a temperature of $T=303\,\mbox{K}$ shown in 
Fig. \ref{fig:vH}
and given in Table \ref{tab:diffu_vap}.

Both the data computed from the KPL and from 
the NGKPL force field as
a function of alkyl chain-length 
are rather close to the experimental data of of Verevkin {\em et al.} \cite{Verevkin:2013}.
\begin{table}
    \centering
    \setlength{\tabcolsep}{0.45cm}
    \caption{MD simulated self-diffusion coefficients of the
[NTf$_{2}$] anion $D_{-}$ as well as vaporization enthalpies
$\Delta_{\rm v}H$ as a function of the alkyl chain-length $n$ in
[C$_{n}$MIm][NTf$_{2}$] for the KPL and the new NGKPL force field. See also Fig. \ref{fig:diffusion_2} and Fig. \ref{fig:vH}.}
    \begin{tabular}{crrrr}\hline\hline
        & \multicolumn{2}{c}{$D_{-}$ / 10$^{-11}$ m$^{2}$\,s$^{-1}$} &
\multicolumn{2}{c}{$\Delta_{\rm v}H$ / kJ\,mol$^{-1}$}\\
    $n$    & KPL    & NGKPL    &    KPL    & NGKPL \\ \hline
    1 & 5.21 & 3.87 & 121.6 & 119.9 \\
    2 & 4.76 & 3.56 & 122.6 & 119.4 \\
    4 & 3.22 & 2.42 & 128.5 & 126.0 \\
    6 & 1.83 & 1.26 & 137.5 & 135.5 \\
    8 & 1.00 & 0.67 & 148.5 & 146.9 \\ \hline\hline
    \end{tabular}
    \label{tab:diffu_vap}
\end{table}
However, we would like to point out, that the optimized NGKPL force field is in even 
better  agreement with the QCM experiments,
particularly for chain-lengths up to $n=4$.
Not only are the data for $n=2$ now in quantitative agreement with the experimental data,
but also the step from $n=1$ to $n=2$ is better captured by the new model, suggesting
a significant influence of the enhanced conformational diversity of the
[NTf$_{2}$] anion \cite{Fumino:2010}. Since the exact slope of $\Delta_{\rm v}H$
as a function of the alkyl chain-length has been shown to be controlled by
the counterbalance of electrostatic and {\sc van der Waals} forces
\cite{Koddermann:2008}, the increasing deviation for longer
chain-length might indicate a slight misrepresentation of size of 
the dispersion interaction introduced by increasing the alkyl chain-length.

\subsection{Viscosities \& Reorientational Correlation Times}

To further compare dynamical properties of the 
simulated ionic liquids with experimental data, 
the temperature dependence of the
reorientational correlation times for the C(2)-H vector
and viscosities for [C$_2$MIm][NTf$_2$] where calculated.
To compare with the quadrupolar relaxation experiments
of Wulf {\em et al.} \cite{Wulf:2007}
we computed reorientational correlation functions $R(t)$ of the C(2)-H bond-vector
according to
\begin{equation}
R(t) = \left< P_2\{\cos[\theta_\mathrm{CH}(t)]\} \right>,
\end{equation}
where $P_2$ is the second {\sc Legendre} polynomial and
\begin{equation}
\cos[\theta_\mathrm{CH}(t)] =
\frac{\vec{r}_\mathrm{CH}(0) \cdot\vec{r}_\mathrm{CH}(t)}{|\vec{r}_\mathrm{CH}|^2}
\end{equation}
represents the angle-cosine between the CH-bond vector at times ``0'' and
$t$ and $|\vec{r}_\mathrm{CH}|$ is the CH-bond length, which is kept fixed during the simulation. The reorientational correlation times $\tau_\mathrm{c}$ are obtained
as integral over the correlation function
\begin{equation}
\tau_\mathrm{c} = \int\limits_0^\infty R(t)\,dt\,.
\end{equation}
Here, the long-time behavior is fitted to a  stretched exponential function
and the total correlation time is determined by numerical integration.
\begin{figure}
	\includegraphics[width=7.0cm]{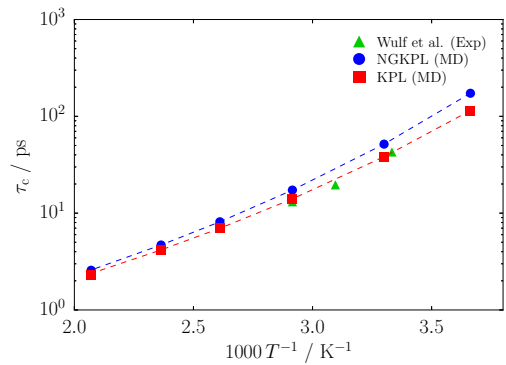}
	\caption{Reorientational correlation time of the C(2)-H vector
	  in [C$_{2}$MIm][NTf$_{2}$] as function of temperature. 
	  The experimental data of Wulf et
al. \cite{Wulf:2007} are shown as green triangles, KPL-data as red squares, 
and NGKPL-data as blue dots. The data are summarized in  Table \ref{tab:visc_tau_c2mim}.}
	\label{fig:tau}
\end{figure}
Again we find that both force fields are in good agreement with the experimental values, 
albeit with the original KPL model being slightly closer to
the experimental data (see Fig. \ref{fig:tau}). 

To determine the viscosities we used the approach of Zhang {\em et al.} \cite{Maginn:2015} 
to compute viscosities from equilibrium-fluctuations of the off-diagonal
elements of the pressure tensor  
via the {\sc Green-Kubo} relation
\begin{align}
	\eta = \frac{V}{k_{\rm B}T} \int^{\infty}_{0} \left< P_{\alpha\beta}(0)\cdot P_{\alpha\beta}(t)\right> dt.
\end{align}
For each temperature we performed $15$ independent $NVT$ simulations, where the starting configurations where sampled  from the earlier $NpT$ simulations with a constant
time interval of $2\,\mbox{ns}$. 
After a $1\,\mbox{ns}$ equlibration we computed $8\,\mbox{ns}$ long productions runs
for each of the sampled configurations storing the pressure tensor 
data for each time-step. Finally, the correlation function was
calculated and integrated over a time-window of $1\,\mbox{ns}$ for each of the $15$ simulations. The average of the running integrals was
calculated as well as standard deviation. The average over the 
running integrals as well as the standard deviation where handled as suggested
by Zhang {\em et al.} \cite{Maginn:2015} with a fitting cut off $t_{\rm cut}$ at 
the point where $\sigma(t)$ is $40\,\%$ of the calculated average viscosity.

\begin{figure}
	\includegraphics[width=7.0cm]{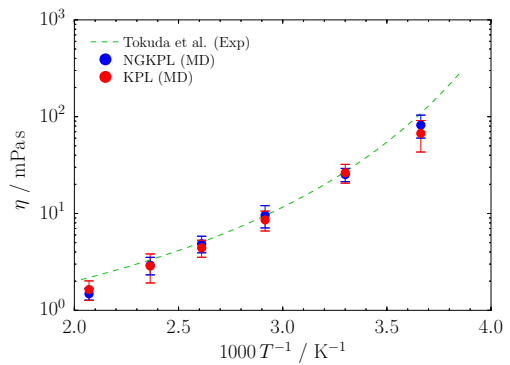}
	\caption{Viscosities as function of temperature for [C$_{\rm 2}$MIm][NTf$_2$] for NGKPL (blue dots) and KPL (red squares).
 The experimental data was taken from Tokuda {\em et al.} (green dashed line) \cite{Tokuda:2005}. See also Table \ref{tab:visc_tau_c2mim}.}
	\label{fig:vis}
\end{figure}

We find that the differences between the KPL and NGKPL models to be rather
small. Both are basically lying within the statistical 
errors of this method. However, both force field model yield
viscosities very close to the experiment (Fig. \ref{fig:vis}).

\begin{table}
     \centering
     \setlength{\tabcolsep}{0.28cm}
     \caption{Viscosities $\eta$ and reorientational correlation times of
the C(2)-H vector $\tau_{\rm c}$ as a function of temperature calculated
from MD simulations of [C$_{2}$MIm][NTf$_{2}$] employing the KPL and the
NGKPL force fields. See also Fig. \ref{fig:tau} and Fig. \ref{fig:vis}.}
     \begin{tabular}{crrrr}\hline\hline
         & \multicolumn{2}{c}{$\eta$ / mPa\,s} &
\multicolumn{2}{c}{$\tau_{\rm c}$ / ps} \\
         $T$ / K & KPL    & NGKPL    &    KPL    & NGKPL \\ \hline
         273 & 67 $\pm$ 24 & 82 $\pm$ 22        &  173.1 & 114.7\\
         303 & 26 $\pm$  6  & 25 $\pm$ 4        &    51.6 &  38.2\\
         343 & 8.6 $\pm$ 2.0 & 9.6 $\pm$ 2.4    &    17.3 &  14.1\\
         383 & 4.4 $\pm$ 0.9 & 4.9 $\pm$ 0.9    &     8.1 &  7.0    \\
         423 & 2.9 $\pm$ 1.0 & 2.9 $\pm$ 0.6    &     4.7 &      4.1\\
         483 & 1.6 $\pm$ 0.3 & 1.48 $\pm$ 0.21    &  2.6 &      2.3\\ \hline\hline
     \end{tabular}
     \label{tab:visc_tau_c2mim}
\end{table}

\section{Conclusions}

We showed that the reparametrization of the dihedral potentials as well as charges of the [NTf$_2$] anion leads to an improvment of the
force field model of K\"oddermann {\em et al.} 
for imidazolium based ionic liquids from 2007. 
The most prominent advantage of the new parameter set
 is that the minimum energy conformations ({\em trans} and {\em gauche}) of the anion, 
 as demonstrated from {\em ab initio} calculations 
 and {\sc Raman} experiments, are now well reproduced.

The results obtained for [C$_n$MIm][NTf$_2$] show that this correction leads to a slightly better agreement between experiment and molecular dynamics
simulation for a variety of properties, such as  densities, diffusion coefficients, vaporization enthalpies, reorientational correlation times, and viscosities. 
Even though we focused on optimizing the anion
parameters, the alkyl chain-length dependence is found to be general also 
closer to the experiment.

With this work we want to point out that it is important to re-examine
 established force field and, if necessary, to improve those. We highly
recommend to use the new NGKPL force field
 for the [NTf$_2$] anion instead of the original 
 KPL force field. Especially for simulation aiming to
describe the thermodynamics, dynamics and 
also structure of imidazolium based ionic liquids.

\section{Acknowlegements}

B.G. is thankful for 
financial support provided by COST Action CM 1206 (ÒEXIL - Exchange on Ionic LiquidsÓ). 


\end{document}